\documentclass[twoside]{article}
\usepackage[T1]{fontenc}
\usepackage{graphicx}
\usepackage{amsmath}

\catcode`\@=11
\long\def\@makefntext#1{
\protect\noindent \hbox to 3.2pt {\hskip-.9pt
$^{{\eightrm\@thefnmark}}$\hfil}#1\hfill}               

\def\@makefnmark{\hbox to 0pt{$^{\@thefnmark}$\hss}}    

\def\ps@myheadings{\let\@mkboth\@gobbletwo
\def\@oddhead{\hbox{}
\rightmark\hfil\eightrm\thepage}
\def\@oddfoot{}\def\@evenhead{\eightrm\thepage\hfil
\leftmark\hbox{}}\def\@evenfoot{}
\def\sectionmark##1{}\def\subsectionmark##1{}}

\oddsidemargin=\evensidemargin
\newcounter{sectionc}\newcounter{subsectionc}\newcounter{subsubsectionc}
\renewcommand{\section}[1] {\vspace{12pt}\addtocounter{sectionc}{1}
\setcounter{subsectionc}{0}\setcounter{subsubsectionc}{0}\noindent
	{\tenbf\thesectionc. #1}\par\vspace{5pt}}
\renewcommand{\subsection}[1] {\vspace{12pt}\addtocounter{subsectionc}{1}
	\setcounter{subsubsectionc}{0}\noindent
	{\bf\thesectionc.\thesubsectionc. {\kern1pt \bfit #1}}\par\vspace{5pt}}
\renewcommand{\subsubsection}[1] {\vspace{12pt}\addtocounter{subsubsectionc}{1}
	\noindent{\tenrm\thesectionc.\thesubsectionc.\thesubsubsectionc.
	{\kern1pt \tenit #1}}\par\vspace{5pt}}
\newcommand{\nonumsection}[1] {\vspace{12pt}\noindent{\tenbf #1}
	\par\vspace{5pt}}

\newcounter{appendixc}
\newcounter{subappendixc}[appendixc]
\newcounter{subsubappendixc}[subappendixc]
\renewcommand{\thesubappendixc}{\Alph{appendixc}.\arabic{subappendixc}}
\renewcommand{\thesubsubappendixc}
	{\Alph{appendixc}.\arabic{subappendixc}.\arabic{subsubappendixc}}

\renewcommand{\appendix}[1] {\vspace{12pt}
	\refstepcounter{appendixc}
	\setcounter{figure}{0}
	\setcounter{table}{0}
	\setcounter{lemma}{0}
	\setcounter{theorem}{0}
	\setcounter{corollary}{0}
	\setcounter{definition}{0}
	\setcounter{equation}{0}
	\renewcommand{\thefigure}{\Alph{appendixc}.\arabic{figure}}
	\renewcommand{\thetable}{\Alph{appendixc}.\arabic{table}}
	\renewcommand{\theappendixc}{\Alph{appendixc}}
	\renewcommand{\thelemma}{\Alph{appendixc}.\arabic{lemma}}
	\renewcommand{\thetheorem}{\Alph{appendixc}.\arabic{theorem}}
	\renewcommand{\thedefinition}{\Alph{appendixc}.\arabic{definition}}
	\renewcommand{\thecorollary}{\Alph{appendixc}.\arabic{corollary}}
	\renewcommand{\theequation}{\Alph{appendixc}.\arabic{equation}}
	\noindent{\tenbf Appendix \theappendixc #1}\par\vspace{5pt}}
\newcommand{\subappendix}[1] {\vspace{12pt}
	\refstepcounter{subappendixc}
	\noindent{\bf Appendix \thesubappendixc. {\kern1pt \bfit #1}}
	\par\vspace{5pt}}
\newcommand{\subsubappendix}[1] {\vspace{12pt}
	\refstepcounter{subsubappendixc}
	\noindent{\rm Appendix \thesubsubappendixc. {\kern1pt \tenit #1}}
	\par\vspace{5pt}}

\newcommand{\textlineskip}{\baselineskip=13pt}
\newcommand{\smalllineskip}{\baselineskip=10pt}

\def\eightcirc{
\begin{picture}(0,0)
\put(4.4,1.8){\circle{6.5}}
\end{picture}}
\def\eightcopyright{\eightcirc\kern2.7pt\hbox{\eightrm c}}

\newcommand{\copyrightheading}[1]
	{\vspace*{-2.5cm}\smalllineskip{\flushleft
	{\footnotesize International Journal of Modern Physics C, #1}\\
	{\footnotesize $\eightcopyright$\, World Scientific Publishing
	 Company}\\
	 }}

\newcommand{\publisher}[2]{{\begin{center}\footnotesize\smalllineskip
	Received #1\\
	Revised #2
	\end{center}
	}}

\def\abstracts#1#2#3{{
	\centering{\begin{minipage}{4.5in}\baselineskip=10pt\footnotesize
	\parindent=0pt #1\par
	\parindent=15pt #2\par
	\parindent=15pt #3
	\end{minipage}}\par}}

\def\keywords#1{{
	\centering{\begin{minipage}{4.5in}\baselineskip=10pt\footnotesize
	{\footnotesize\it Keywords}\/: #1
	\end{minipage}}\par}}

\renewenvironment{thebibliography}[1]
	{\frenchspacing
	 \ninerm\baselineskip=11pt
	 \begin{list}{\arabic{enumi}.}
	{\usecounter{enumi}\setlength{\parsep}{0pt}
	 \setlength{\leftmargin 12.7pt}{\rightmargin 0pt} 
	 \setlength{\itemsep}{0pt} \settowidth
	{\labelwidth}{#1.}\sloppy}}{\end{list}}

\newcounter{itemlistc}
\newcounter{romanlistc}
\newcounter{alphlistc}
\newcounter{arabiclistc}

\newcommand{\fcaption}[1]{
	\refstepcounter{figure}
	\setbox\@tempboxa = \hbox{\footnotesize Fig.~\thefigure. #1}
	\ifdim \wd\@tempboxa > 5in
	   {\begin{center}
	\parbox{5in}{\footnotesize\smalllineskip Fig.~\thefigure. #1}
	    \end{center}}
	\else
	     {\begin{center}
	     {\footnotesize Fig.~\thefigure. #1}
	      \end{center}}
	\fi}

\newcommand{\tcaption}[1]{
	\refstepcounter{table}
	\setbox\@tempboxa = \hbox{\footnotesize Table~\thetable. #1}
	\ifdim \wd\@tempboxa > 5in
	   {\begin{center}
	\parbox{5in}{\footnotesize\smalllineskip Table~\thetable. #1}
	    \end{center}}
	\else
	     {\begin{center}
	     {\footnotesize Table~\thetable. #1}
	      \end{center}}
	\fi}

\def\@citex[#1]#2{\if@filesw\immediate\write\@auxout
	{\string\citation{#2}}\fi
\def\@citea{}\@cite{\@for\@citeb:=#2\do
	{\@citea\def\@citea{,}\@ifundefined
	{b@\@citeb}{{\bf ?}\@warning
	{Citation `\@citeb' on page \thepage \space undefined}}
	{\csname b@\@citeb\endcsname}}}{#1}}

\newif\if@cghi
\def\cite{\@cghitrue\@ifnextchar [{\@tempswatrue
	\@citex}{\@tempswafalse\@citex[]}}
\def\citelow{\@cghifalse\@ifnextchar [{\@tempswatrue
	\@citex}{\@tempswafalse\@citex[]}}
\def\@cite#1#2{{$\null^{#1}$\if@tempswa\typeout
	{IJCGA warning: optional citation argument
	ignored: `#2'} \fi}}

\def\pmb#1{\setbox0=\hbox{#1}
	\kern-.025em\copy0\kern-\wd0
	\kern.05em\copy0\kern-\wd0
	\kern-.025em\raise.0433em\box0}

\def\fnt#1#2{\footnotetext{\kern-.3em
	{$^{\mbox{\scriptsize #1}}$}{#2}}}

\def\fpage#1{\begingroup
\voffset=.3in
\thispagestyle{empty}\begin{table}[b]\centerline{\footnotesize #1}
	\end{table}\endgroup}

\def\runninghead#1#2{\pagestyle{myheadings}
\markboth{{\protect\footnotesize\it{\quad #1}}\hfill}
{\hfill{\protect\footnotesize\it{#2\quad}}}}
\font\tenrm=cmr10
\font\tenit=cmti10
\font\tenbf=cmbx10
\font\bfit=cmbxti10 at 10pt
\font\ninerm=cmr9

\font\eightrm=cmr8

\input{psfig.sty}
\usepackage{amsmath}

\begin{document}

\runninghead{M.~Nitti, A.~Torcini \& S.~Ruffo}{An Integration Scheme
for Reaction-Diffusion Models}
\normalsize\textlineskip
\thispagestyle{empty}
\setcounter{page}{1}
\copyrightheading {Vol. 0, No. 0 (1999) 000--000}
\vspace*{0.88truein}
\fpage{1}

\centerline{\bf AN INTEGRATION SCHEME FOR REACTION-DIFFUSION MODELS}
\vspace*{0.37truein}
\centerline{\footnotesize MASSIMO~NITTI}
\vspace*{0.015truein}
\centerline{\footnotesize\it Dipartimento di Sistemi e Informatica,
via S. Marta 3, I-50139 Firenze, Italy}
\centerline{\footnotesize\it E-mail: massimo@lyapu.dsi.unifi.it}
\vspace*{0.15truein}
\vspace*{0.15truein}
\centerline{\footnotesize ALESSANDRO~TORCINI}
\vspace*{0.015truein}
\centerline{\footnotesize\it INFM, Udr Firenze,
L.go E. Fermi 5, I-50125 Firenze, Italy}
\centerline{\footnotesize\it Dipartimento di Energetica,
via S. Marta 3, I-50139 Firenze, Italy}
\centerline{\footnotesize\it E-mail: torcini@fi.infn.it}
\vspace*{0.30truein}

\centerline{\footnotesize STEFANO~RUFFO}
\vspace*{0.015truein}
\centerline{\footnotesize\it Dipartimento di Energetica,
via S. Marta 3, I-50139 Firenze, Italy}
\centerline{\footnotesize\it INFN, sez. Firenze}
\centerline{\footnotesize\it E-mail: ruffo@fi.infn.it}

\vspace*{0.225truein}
\publisher{(May 07,1999)}{(May 12,1999)}
\vspace*{0.21truein}

\abstracts{
A detailed description and validation of a recently developed
integration scheme is here reported for one- and two-dimensional
reaction-diffusion models. As paradigmatic examples of this
class of partial differential equations the complex Ginzburg-Landau
and the Fitzhugh-Nagumo equations have been analyzed.
The novel algorithm has precision and stability comparable
to those of pseudo-spectral codes, but it is more convenient to employ
for systems with quite large linear extention $L$.
As for finite-difference methods, the implementation of the present 
scheme requires only information about the local enviroment 
and this allows to treat also system with very complicated 
boundary conditions.
}{}{}
\vspace*{10pt}
\keywords{Partial Differential Equations; Reaction-Diffusion Models;
Integration Schemes.
}
\vspace*{1pt}\textlineskip

\section{Introduction\label{sec:Intro}}

The study and development of efficient and accurate integration 
algorithms for reaction-diffusion partial differential 
equations (RDPDE) is a compelling task. Because reaction-diffusion 
equations are used to represent quite a large class of systems 
ranging from chemical,
to biological and physical ones \cite{bettel}. Within the field of physics
deterministic reaction-diffusion models have been used to 
mimick front-propagation, 
surface- and interface-growths, turbulent behaviour, pattern
formation, excitable media, etc. \cite{cross,grinrod,vans}.

A paradigmatic example of RDPDE
is the well-known Fitzhugh-Nagumo equation (FHNE)
used to reproduce the onset of solitary waves, periodic wave
trains, circular and spiral waves in excitable media
\cite{tyson,winfree}
\begin{equation}
\label{fhn}
u_t = D \nabla^2 u + u(u-a)(1-u) -v \quad ; \quad
v_t = \varepsilon (\beta u - v)
\end{equation}
where $u=u({\bf r},t)$ and 
$v=v({\bf r},t)$ are real fields and 
represent the activator and the inhibitor,
respectively. The appearance of different
dynamical behaviours and of the associated
patterns is ruled by the values assumed
by the real parameters $D$, \, $a$, \, $\varepsilon$ and $\beta$ 
and by the dimensionality of the system
(see Refs. \cite{tyson,winfree} for a review).

Another widely studied  reaction diffusion model
is represented by the complex Ginzburg-Landau equation (CGLE),
that is particularly 
relevant for the description of the dynamics of
spatially extended systems both
from an experimental and from a theoretical point of 
view. This because any spatially extended systems
that undergoes a Hopf bifurcation from a stationary
to an oscillatory state is described by the CGLE
sufficiently close to the bifurcation point \cite{kuramoto}.
The CGLE can be written as
\begin{equation}
\label{cgle}
A_t = (1+j c_1) \nabla^2 A + A + (1-j c_3) |A|^2 A
\end{equation}
where $c_1$ and $c_3$ are real positive numbers, while
$A({\bf r},t) = \rho({\bf r},t) \exp{[j \psi({\bf r},t)]}$
is a complex field of amplitude $\rho$ and phase $\psi$.
Here $j$ is the imaginary unit with $j^2 = -1$.
Equation (\ref{cgle}) exhibits several different stationary
and turbulent regimes and a quite rich literature has been
devoted to the description of its phases \cite{shra,tor2}.
In the limit $(c_1,c_3) \to (\infty,\infty)$ the CGLE
reduces to the nonlinear Schr{\"o}dinger equation,
while the real Ginzburg-Landau equation is recovered
in the opposite limit $(c_1,c_3) \to (0,0)$ 
\cite{kuramoto}.

In the present paper we describe in detail the
implementation and the performances of a new integration 
scheme for RDPDE. In particular, we deal with the
FHNE and the CGLE in one and two dimensions for different types
of boundary conditions.
In Section II the new integration algorithm is explained
and its application to the FHNE and to the CGLE is described
in detail. The treatment of different boundary conditions
(namely, periodic, no-flux and fixed ones) is explained
in Section III. In Sect. IV the precision and the performances 
of the
present scheme are compared with those of usual pseudo-spectral
codes for one- and two-dimensional systems.
A brief final discussion is reported in Sect. V.

\section{The integration scheme\label{sec:scheme}}

The algorithm here reported is a modification of 
the well known "leap-frog" method applied to a spatially
extended system, whose evolution is represented by a partial 
differential equation (PDE). The leap-frog algorithm was originally
devised for Hamiltonian systems \cite{verlet} and
heavily employed for classical molecular dynamics
simulations \cite{allen}. But it can
be also implemented for any ordinary differential equation,
when the time evolution of a given dynamical variable
${\bf z} ={\bf z} (t)$ 
can be written as
\begin{equation}
\label{eq1}
{\bf \dot z}(t) = A \enskip {\bf z}(t) + B \enskip {\bf z}(t)
\end{equation}
where $A$ and $B$ are two operators, generically non linear,
that typically do not commute. 
Suppose now that one is able to integrate separately
the two contributions associated to $A$ and $B$ 
in Eq. (\ref{eq1}), but not Eq. (\ref{eq1}) as a whole.
Therefore one would like to rewrite the formal solution 
of (\ref{eq1}) ${\bf z(t)} =\exp{[t(A+B)]} {\bf z(0)}$ 
as a product of terms $\exp{[a_k t A]}$ and
$\exp{[b_k t B]}$. A vast literature has been devoted to
the choice of the coefficients $a_k$ and $b_k$ that minimizes
the errors done in rewriting the formal solutions in such
an approximate way \cite{yoshida,atela}. The best
known approximation is the so-called Trotter formula
\begin{equation}
\label{trotter}
{\rm e}^{\tau(A+B)} = 
{\rm e}^{\tau A/2} {\rm e}^{\tau B}{\rm e}^{\tau A/2} + o(\tau^3)
\quad ,
\end{equation}
where $\tau$ is the time step.
For a time interval $t=\tau n$, one can combine more efficiently
$n-1$ half steps and obtains
\begin{equation}
\label{trotter2}
{\rm e}^{t(A+B)} \simeq
{\rm e}^{\tau A/2} \left[
{\rm e}^{\tau B} {\rm e}^{\tau A} \right]^{(n-1)}
{\rm e}^{\tau B} {\rm e}^{\tau A/2} 
\end{equation}
which essentially corresponds to the 
leap-frog (or Verlet) algorithm.

\noindent
Let us now consider a generic reaction diffusion
equation
\begin{equation}
\label{eqr}
{\bf \dot V} =  {\bf G}({\bf V}) + D \nabla^2
{\bf V}
\end{equation}
where ${\bf V} = {\bf V}({\bf x},t)$ is a classical field.
The most used techniques to integrate such a partial differential
equation (PDE) are the so-called time-splitting pseudo-spectral codes
\cite{press}. These schemes are among the most stable and efficient
and we will illustrate it in the case of the leap-frog time 
splitting (\ref{trotter2})\cite{frauenk}. 
Referring to (\ref{eq1}), the operators
are identified in the following manner
\begin{equation}
\label{operator}
A {\bf V} = {\bf G}({\bf V}) \quad , \quad
B {\bf V} =  D \nabla^2 {\bf V} 
\end{equation}
where now $A$ is a non-linear local (in space) operator
and $B$ a linear non-local one. One should now solve
separately the evolution equation involving the operator
$A$ and $B$ and then apply formula (\ref{trotter2}).
The solution of the non-linear part is usually performed
by standard integration techniques for ODE, but in some
cases it can be even solved exactly (as it will be shown for the
CGLE). To integrate the part involving the spatial
derivatives, we consider the Fourier transform of the
field ${\bf \tilde V} ({\bf p},t) $ and of 
the corresponding evolution equation
\begin{equation}
\label{fft}
{\bf \dot {\tilde V}} =  - p^2 D 
{\bf \tilde V}
\end{equation}
whose solution is simply 
\begin{equation}
\label{fft2}
{\bf {\tilde V}}
({\bf p},t+\tau) 
= \exp{[-p^2 D \tau]} {\bf \tilde V}
({\bf p},t) 
\end{equation}
In order to obtain the solution in the real space it
is now sufficient to back Fourier transform the signal.
Therefore, an integration over a time $\Delta t$ with discrete
time step $\tau =\Delta t/n$ and adopting a leap-frog scheme can
be summarized as
\begin{equation}
\label{fft3}
{\bf  V}({\bf r},t +\Delta t)=  
\left\{
{\rm e}^{\frac{\tau}{2}A} \left[ 
F^{-1} {\rm e}^{- p^2 D \tau}
F {\rm e}^{\tau A} \right]^{(n-1)}
F^{-1} {\rm e}^{- p^2 D \tau} 
F {\rm e}^{\frac{\tau}{2} A} 
\right\}
{\bf  V}({\bf x},t)  
\end{equation}
where $F$ indicates the Fourier transform and $F^{-1}$
the reverse transform.

\noindent
The integration method we propose follows exactly the
same scheme as the one just shown, but the step involving 
spatial derivatives is directly solved in the real space.
As previously noticed, integration in Fourier
space reduces to a simple product (see (\ref{fft2}));
in the direct space this corresponds to perform a 
convolution integral of the signal with an appropriate kernel
\begin{equation}
\label{conv}
{\bf  V}
({\bf r},t + \tau) 
= \int d{\bf s}  K({\bf s},\tau) {\bf V} 
({\bf r - s},t) \quad ,
\end{equation}
where ${\bf r}=(x_1,x_2,x_3)$ in the three dimensional case.
The expression of the kernel in 3d for the 
reaction-diffusion systems is straightforward 
\begin{equation}
\label{kernel}
K({\bf r},t) = 
\frac{1}{(4 D \pi t)^{3/2}} 
\exp{\left[ \frac{-|r|^2}{4Dt} \right]} \quad .
\end{equation}

In order to implement this scheme on a computer, we have to
deal with a discrete time step $\tau$
and with a discrete spatial resolution ${\bf \Delta}
=(\Delta_1,\Delta_2,\Delta_3)$. The discrete
version of the convolution integral  (\ref{conv})
reads
\begin{equation}
\label{conv_dis}
{\bf  V} ({\bf l},i+1) 
= \sum_{n_1,n_2,n_3} K({\bf n}) {\bf V}
({\bf l - n},i) 
\end{equation}
where ${\bf r}= (\Delta_1 n_1,\Delta_2 n_2,\Delta_3 n_3)$ and
${\bf n}=(n_1,n_2,n_3)$  is the spatial index, while $i$ is the temporal one 
($t = i \tau$) and the sum
runs over all the lattice. This approach 
maintains exactly the same accuracy of the spectral
codes, but it is inefficient in terms
of CPU time. Since the kernel decays rapidly along
any spatial direction, a first naive improvement could consist in 
evaluating the sum in (\ref{conv_dis}) only over $q$ nearest neighbours
of each considered site, but this approach, for precisions comparable
to those of pseudo-spectral codes, is also quite time 
consuming \cite{tor1}.

\noindent
A better strategy consists in truncating the sum
in Eq. (\ref{conv_dis}) and in substituting 
the discretized kernel with a set of unknown coefficients
$C({\bf n})$, that should be conveniently determined 
\cite{tor2}
\begin{equation}
\label{conv_opt}
{\bf  V} ({\bf l},i+1) 
= \sum_{n_1,n_2,n_3=-N_c}^{N_c} C({\bf n}) {\bf V} 
({\bf l - n},i) 
\end{equation}

\noindent
where $N_c$ is the number of convolution channels considered
along each spatial direction.
We have devised the following method to determine the 
$(2 N_c +1)^d$ coefficients $C({\bf n})$ for a $d$-dimensional
lattice:

\begin{itemize}

\item
let us first make the {\it ansatz} that the field can 
be well approximated, on the considered grid, by decomposing 
it on $(2 N_c+1)^d$ elements of the following Fourier basis 
$$
\left\{\exp{\left[j \sum_h
k_h \alpha_h x_h \right]} \right\}
$$
where $k_h = -N_c, \dots , N_c$ and $\alpha_h$ is a set
of parameters to be determined. On this basis the field
can be expressed as
$$
{\bf  V} ({\bf l},i) 
= \sum_{\{ k_h \}}
{\bf \tilde V} (\{ k_h \},i) 
\enskip
\exp{\left[j \sum_{h=1}^d
k_h \alpha_h l_h \Delta_h \right]} \quad ;
$$

\item
the time evolution of each 
Fourier components is determined
by the time propagator (\ref{fft})
$$
\exp{\left[-\sum (k_h \alpha_h)^2 D \tau \right]} \quad ;
$$

\item 
substituting in eq. (\ref{conv_dis}) the field decomposed
on the Fourier basis and imposing that the kernel should
give the exact time evolution on these modes, one is left
with the following set of equations :
\begin{equation}
\label{sys}
\sum_{\bf n}  \exp{\left[ - j \sum_{h=1}^d
k_h \alpha_h n_h \Delta_h \right]} 
C({\bf n}) =
\exp{\left[-(\sum (k_h \alpha_h)^2 D \tau \right]}
\end{equation}
$\forall k_h = -N_c, \dots , N_c \quad ,$
from which one can obtain the coefficients $C({\bf n})$.

\end{itemize}

\noindent
In isotropic problems
the spatial resolution does not depend on the
considered axis (i.e. $\Delta_h=\Delta \enskip \forall
h$)
and in that case symmetry reasons suggest that
$\alpha_h=\alpha \forall h$. Moreover, the
symmetry of the kernel allows to reduce the 
$(2 N_c +1)^d$ system
to a system of $\left\{(N_c+d)!/[N_c!d!]-1\right\}$ equations. 
This because the elements
of the kernel are related by the following symmetry 
relations
\begin{equation}
C(n_1,n_2,n_3) = C^\prime([n^2_1+n^2_2+n^2_3])
\end{equation}
and by the normalizations condition
\begin{equation}
C(0,0,0) = 1 - \sum_{n_1,n_2,n_3 \ne (0,0,0)}
C(n_1,n_2,n_3) 
\end{equation}

\begin{figure}[htb]
\centerline{\psfig{figure=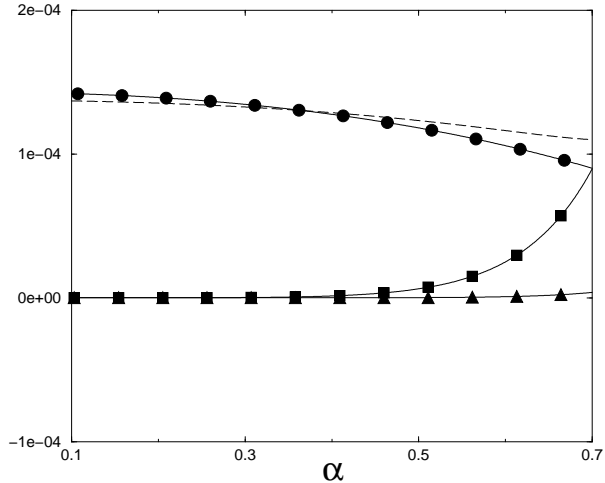,width=8cm,angle=270}}
\caption{\label{fig:alfa}
Dependence of the integration error (dashed line) and
of the $SSQ$ (reported in arbitrary units)  (circles) 
on the choice of the
parameter $\alpha$ for the one-dimensional FHNE, with $\Delta_x = 0.5$,
$\tau = 0.10$, $N_c = 5$ and no-flux boundary conditions.
The differences between the values of the $4^{th}$- (triangles)
and $6^{th}$-momenta (squares) of the corresponding kernel
and those of the true one (\ref{kernel}) are also reported.
Numerical instabilities have been usually found for
small $\alpha$-values, that in the present case correspond
to $\alpha < 0.08$.
}
\end{figure}

\noindent
However, the problem of the choice of the parameters $\{\alpha_h \}$
is left unsolved. 
Its values are determined by demanding minimization
of the error due to the use of a Fourier basis that is not
the complete one. The $\alpha_h$-parameters should be determined
for each choice of the time step, of the spatial resolution
and of the diffusion coefficient. An empirical strategy to choose
the optimal $\alpha_h$-values, that turns out to be essentially 
correct, consists in finding the values that minimize the quadratic
sum $SSQ = \sum_{n_1,n_2,n_3 = -N_c}^{N_c} | C(n_1,n_2,n_3) |^2$,
with the additional constraint that the first 6 cumulants of the
discretized kernel should coincide (within an arbitrary precision)
with those of the true kernel (\ref{kernel}).
In Fig. \ref{fig:alfa} it is shown that such empirical recipe indeed coincides
with the request of a minimal integration error. 

\section{Specific applications\label{sec:examples}}

\noindent
In this Section we will describe in detail the implementation 
of our integration scheme for the CGLE and the FHNE, 
both in one and two dimensions.

\begin{figure}[htb]
\centerline{\psfig{figure=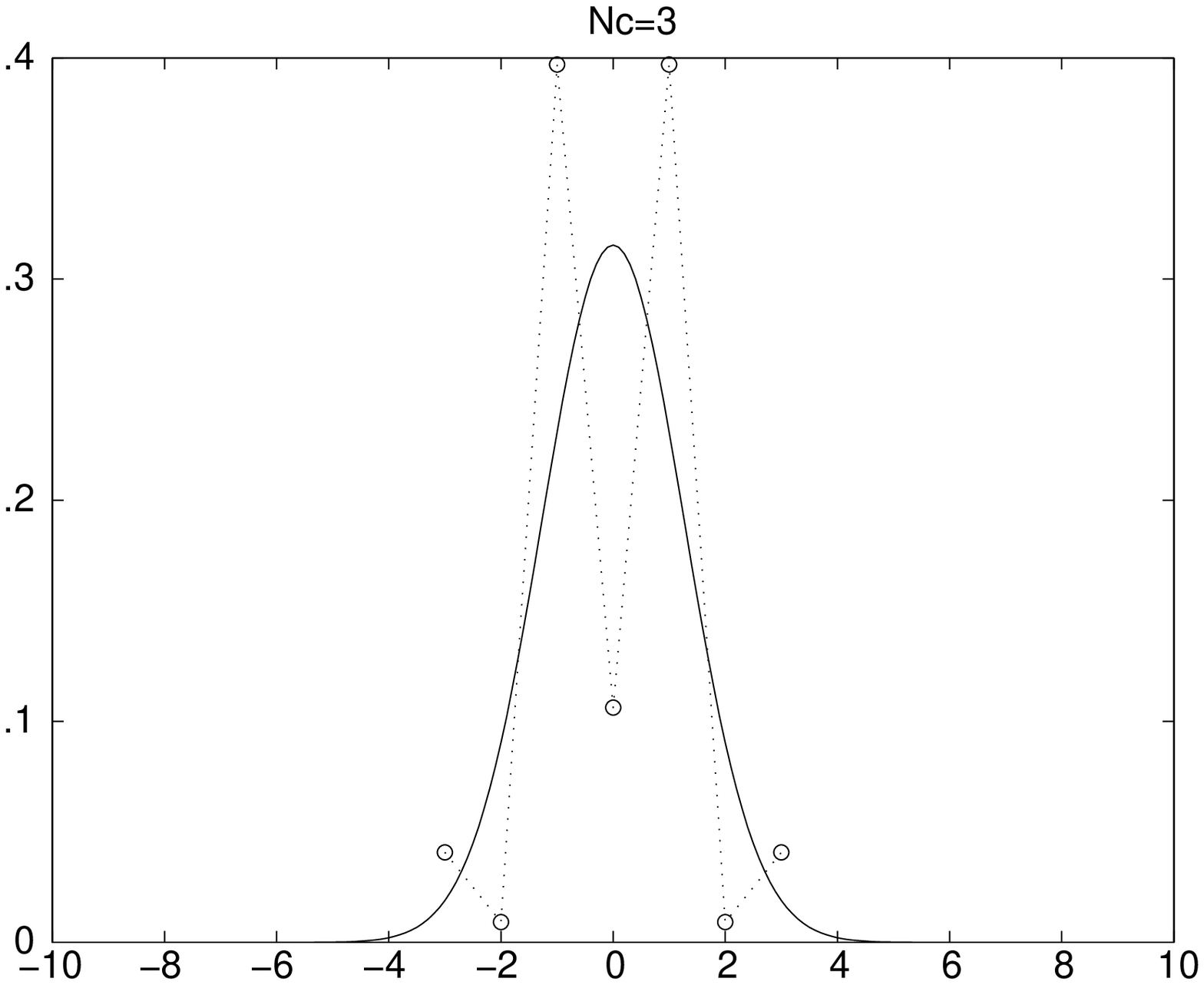,width=6cm}}
\centerline{\psfig{figure=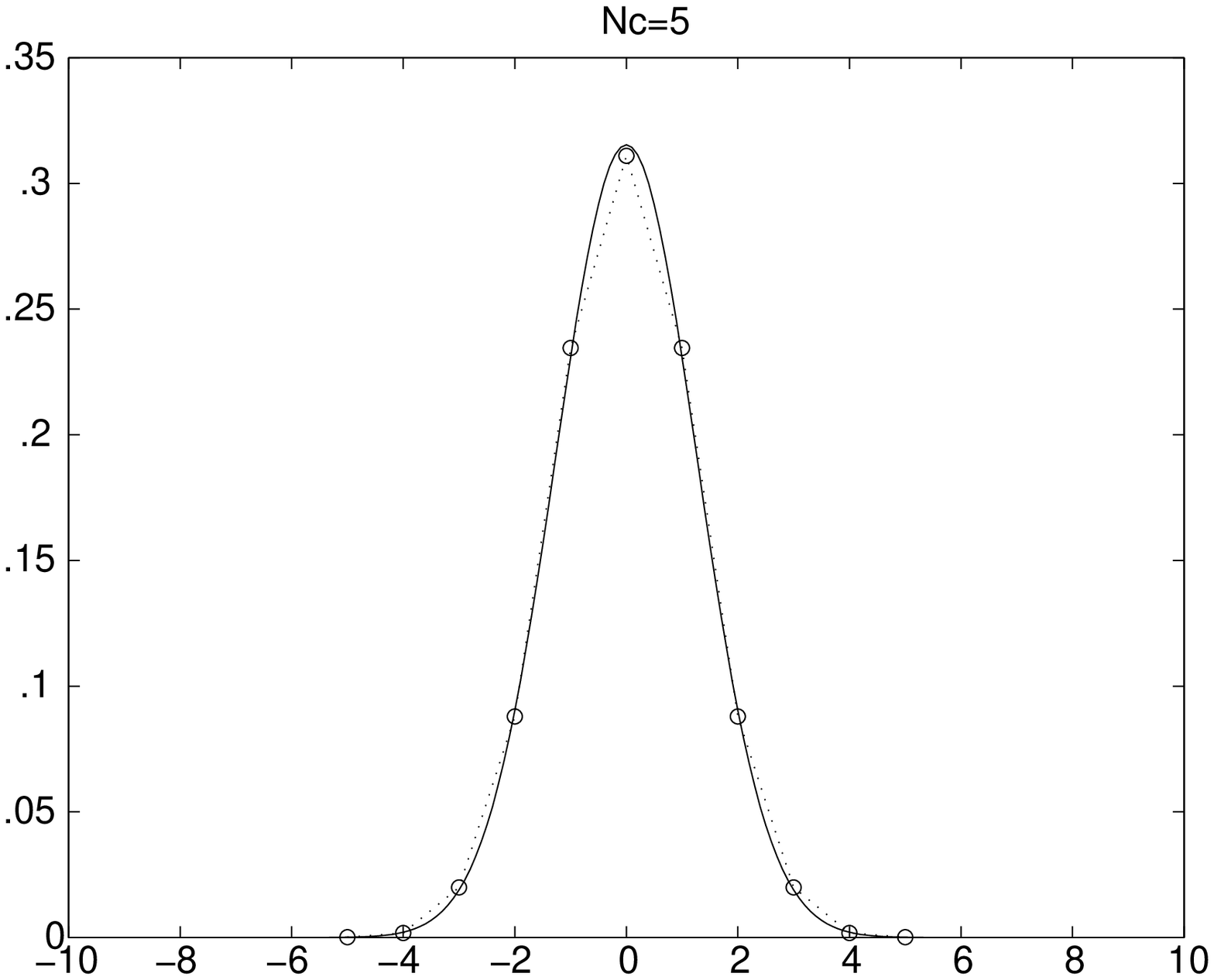,width=6cm}}
\centerline{\psfig{figure=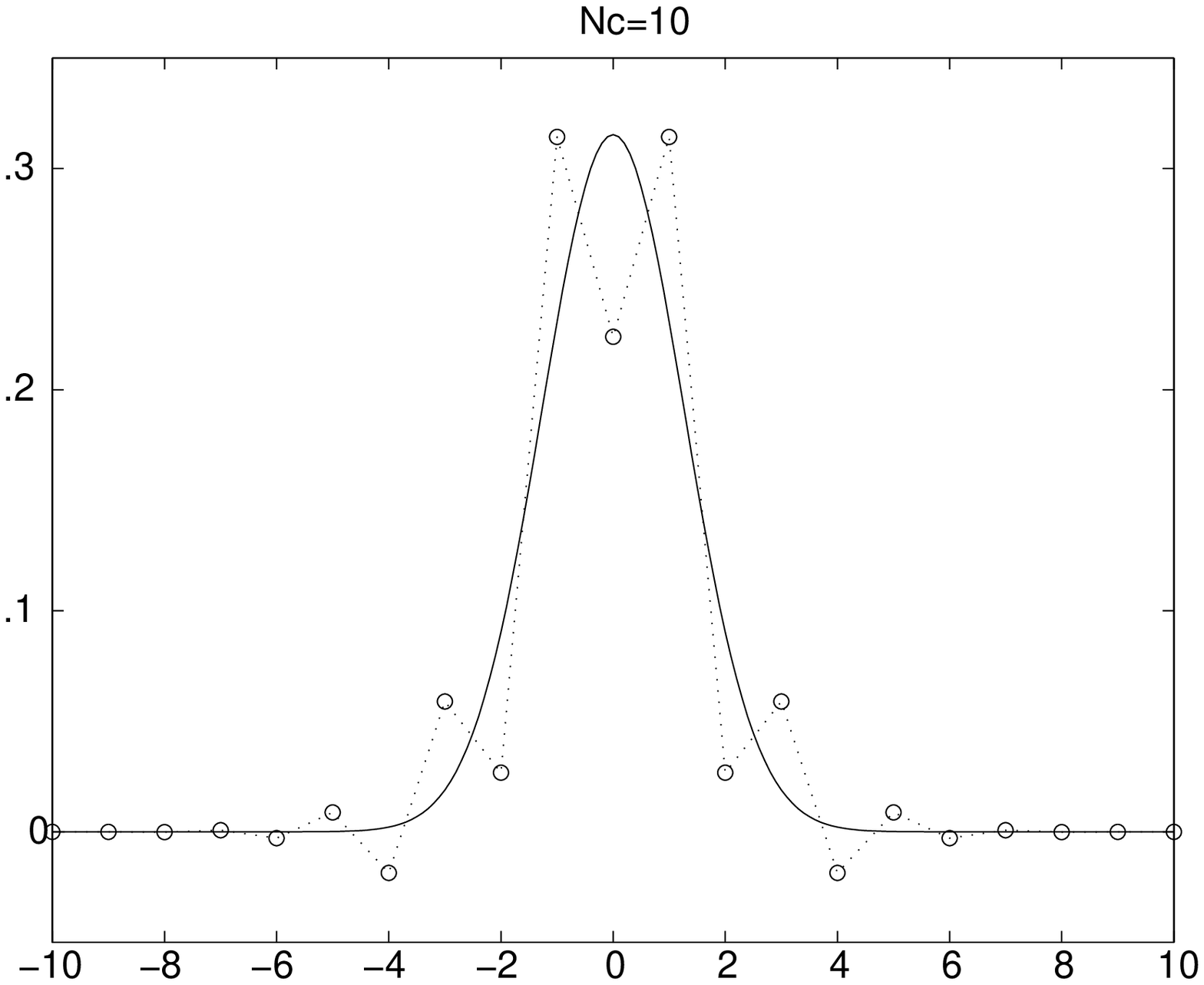,width=6cm}}
\caption{\label{fig:kernel} 
Comparison of the true kernel with the optimized one 
for three different values of $N_c = 3,5,10$
for the one-dimensional FHNE, obtained employing the following parameters
$\tau=0.05$,\, $\Delta_x = 0.25$ and $\alpha = 0.10$.
}
\end{figure}

\noindent
Let us first consider the FHNE, in this case we have chosen
values of the parameter for which the system is in an excitable
state, namely $a = 0.015$, $\beta=1$, $\varepsilon=0.006$
and $D = 1 $. We have performed the integration of
the nonlinear part with a simple Euler scheme,
while for what concerns the diffusive part of the
equation we have simply to integrate the equation:
\begin{equation}
u_t = D \nabla^2 u
\end{equation}
This can be done by adopting our scheme 
where the kernel is given by the real
function reported in (\ref{kernel}). In Fig.~\ref{fig:kernel}
a comparison between the "real" kernel
$K({\bf r})$ and its optimized expression
$C({\bf k})$ on the spatio-temporal grid,
obtained by solving the system (\ref{sys}),
is reported.

\noindent
A more interesting situation is represented by
the CGLE, because this time one is able to solve
exactly the differential equation involving 
the nonlinear operator. This result was
firstly reported in \cite{goldman}, where
the authors have shown that, at least for the CGLE, the precision of 
standard time-splitting pseudo-spectral integration schemes
is superior to that associated with
predictor-corrector and Runge-Kutta algorithms. 
In order to solve such ODE one
has to write separately the following equations for
the phase 
\begin{equation}
\label{fas}
\frac{\partial \psi({\bf r},t)}{\partial t}
= c_3 \rho^2({\bf r},t)
\end{equation}
and for the amplitude
\begin{equation}
\label{amp}
\frac{\partial \rho^2({\bf r},t)}{\partial t}
= 2[ \rho^2({\bf r},t) - \rho^4({\bf r},t)]
\quad .
\end{equation}
The solution of Eq. (\ref{amp}) is simply
given by
\begin{equation}
\label{samp}
\rho({\bf r},\tau)
= 1/
\sqrt{{\rm e}^{-2\tau} (1/\rho^2({\bf r},0)-1)+1}
\end{equation}
and inserting this into eq. (\ref{fas}) one obtains
\begin{equation}
\label{sfas}
\psi({\bf r},\tau)
= \psi({\bf r},0) + c_3 
\left\{
\tau + \ln [\rho({\bf r},0)/\rho({\bf r},\tau)]
\right\} \quad .
\end{equation}

For what concerns the integration of the diffusive
part, it should be noticed that this time the
kernel is complex and it has the expression
\begin{equation}
K({\bf r}) =
\left( \frac{\gamma}{\pi} \right)^{d/2}
{\rm e}^{-\gamma_r |r|^2}\left[
\cos(\gamma_i |r|^2) -j \sin(\gamma_i |r|^2)
\right]
\end{equation}
where 
$\gamma = \gamma_r - j \gamma_i = 1/[4 \tau (1 + j c_1)]$.

\section{Treatment of boundary conditions\label{sec:boundary}}

The most commonly used boundary conditions are:  periodic (P), 
no-flux (NF) (where the first derivative of the field
vanishes at the boundaries) and fixed (F0) (where the
field should have a fixed value at the boundaries,
typically set to zero). To treat these different cases
with pseudo-spectral codes, one is obliged to use
different Fourier basis \cite{press} : namely,
the complex basis for P, the cosine basis for NF and
the sine basis for F0. Instead, with the present algorithm
the same kernel can be used for each of the above cited 
cases and different boundary conditions are treated
just manipulating in different ways the field values
at boundaries. This allows to treat, e.g. a discretized
two-dimensional field with different boundary conditions
on each border or even with various boundary conditions  
on each single border.

Let us consider a classical field 
$\{ {\bf V}(k,i) \}$, with $k =0, \dots, N-1$,
discretized on a one-dimensional grid of resolution $\Delta x$ at a fixed time 
$i$. Problems arise in the evaluation of the convolution
(\ref{conv_opt}) when $k \ge N-N_c-1$ and $k \le N_c$, when
points "outside" the chain should be considered.
For P conditions the field is replicated out of
the chain following the simple rules 
${\bf V}(N+m,i) = {\bf V}(m,i)$ and
${\bf V}(-m,i) = {\bf V}(N-m,i)$. In such
case it should be noticed that the first 
point of the grid is located at any point 
of the chain, while the last point is placed
at a distance $L$, where $N \times \Delta_x = L$.
The NF conditions are treated differently, in this
case we have ${\bf V}(N+m,i) = {\bf V}(N-m+1,i)$ and
${\bf V}(-m,i) = {\bf V}(m+1,i)$, moreover the first
and the last point of the grid are located inside the
chain at a distance $(\Delta_x)/2$ from its extrema. Finally,
the P0 conditions are implemented as follows : 
${\bf V}(N+m,i) = -{\bf V}(N-m,i)$ for $m \ne 0$ and
${\bf V}(N,i) = 0$, while ${\bf V}(-m,i) = -{\bf V}(m,i)$.
In this last situation, the first point of the
grid is located at the beginning of the chain, while
the last point is within the chain at a distance
$\Delta_x$ from the extremum.

\section{Precision and Performances of the Code\label{sec:precision}}

In the present Section we will compare our algorithm 
with standard time splitting codes
\cite{goldman} for the one- and two-dimensional FHNE and CGLE.
It should be stressed that 
the integration algorithm introduced here and the
usual time splitting codes \cite{goldman}
differ only in the treatment of the spatial derivatives.
Therefore for what concerns the analysis of the precision
and of the performances we will concentrate on such
integration step. 

In order to give a measure of the precision
of the considered algorithms,
we have estimated the mean square deviation (MSD)
between an orbit obtained by the algorithm under test
and a reference orbit, that is assumed to be 
"exact". In particular, the MSD has been
evaluated over a fixed time interval $\tau$
and averaged over a total integration time $T = n \tau $. 
Once the test orbit has been periodically synchronized
at the reference one at times $t = m \tau$ $(m=1,\dots,n)$.

In the considered time-split integration schemes we can
identify two different type of errors, one due to
the spatial resolution and the other to the temporal one.
The second kind of error can be measured considering
a test and reference orbit obtained with the same spatial
resolution but with different time integration steps $\tau$
(obviously that of the reference orbit should be taken
much smaller, typically we have considered $\simeq \tau/5$). 
The "temporal" MSD are
obviously identical for our algorithm and for the standard
pseudo-spectral ones. Moreover, with the spatial resolution
considered (that corresponds to typical values employed in literature)
this error is dominant with respect to the "spatial" one.
Data are reported in Table I.

\begin{table}[h]
\begin{center}
\begin{tabular}{c c c c}
\hline
\hline
$\tau$ & 1d & 2d \\
\hline
\hline
0.1  & .164E-03  & .742E-02 \\
0.05  & .741E-04  & .301E-03 \\
\hline
\hline
\end{tabular}
\end{center}
\caption{MSD associated to the temporal integration 
for the FHNE: the data refers to a spatial resolution
$\Delta_x =0.5$ and no-flux boundary conditions.}
\end{table}

For what concerns the MSD associated to the spatial resolution,
it is clear from the data reported in Table II, III, IV and V that
this error is different for the 2 algorithms and it depends strongly
on the number of convolution channels $N_c$ (when our algorithm is
considered). Our algorithm being devised in order to reproduce
the precision of pseudo-spectral codes (that we will 
term FFT), one expects that the MSD associated to our algorithm
will be bigger than that associated to the FFT code for any value
of $N_c$. However, already for $N_c \simeq 20 -30$ in the one- and 
two-dimensional cases the precision of the FFT code is recovered.

\begin{table}[h]
\begin{center}
\begin{tabular}{c c c}
\hline
\hline
$\Delta_x$  & 1.0 & $ 2.0 $\\
\hline
\hline
$N_c=5$  &  2.961E-07  & 9.399E-05  \\
$N_c=10$  & 7.770E-08  & 8.570E-05 \\
$N_c=20$  & 6.593E-08  & 7.289E-05 \\
FFT & 6.727E-08   & 7.513E-05  \\
\hline
\hline
\end{tabular}
\end{center}
\caption{MSD associated to the spatial integration 
relative to our algorithm for various values of $N_c$,
the last row refers to usual pseudo-spectral code.
The data have been obtained for the one-dimensional FHNE
with time step $\tau =0.05$ and no-flux boundary conditions.}
\end{table}

\begin{table}[h]
\begin{center}
\begin{tabular}{c c c}
\hline
\hline
$\Delta_x$  & 1.0 & 2.0 \\
\hline
\hline
$N_c=3  $& 9.115E-07    & 7.619E-05   \\
$N_c=5 $ & 1.256E-07    & 4.913E-05   \\
$N_c=6 $ & 6.884E-08    & 4.264E-05   \\
$N_c=7$  & 5.128E-08    & 4.292E-05   \\
$N_c=8$  & 3.596E-08    & 4.066E-05   \\
$N_c=10$ & 2.838E-08  & 4.102E-05   \\
FFT &1.942E-08    & 2.835E-05 \\
\hline
\hline
\end{tabular}
\end{center}
\caption{MSD associated to the spatial integration 
relative to our algorithm for various values of $N_c$,
the last row refers to usual pseudo-spectral code.
The data have been obtained for the two-dimensional FHNE
with time step $\tau =0.10$ and no-flux boundary conditions.}
\end{table}

\begin{table}[h]
\begin{center}
\begin{tabular}{c c c c}
\hline
\hline
$\Delta_x$ & 2.00 & 1.00 & 0.50\\
\hline
\hline
$N_c = 6$ & 0.125E-04   & 0.266E-06  &   0.123E-08   \\
$N_c=10 $ & 0.955E-05   & 0.187E-07  &  0.120E-11   \\
$N_c=20 $ & 0.227E-05  & 0.109E-08 &    \\
$N_c=30 $ & 0.211E-05  & 0.617E-09 &  0.109E-13 \\
$N_c=40 $ & 0.455E-05  & 0.525E-09 &  0.921E-14 \\
$N_c=50 $ & 0.212E-05 &  0.521E-09 &   0.866E-14 \\
FFT & 0.212E-05 &  0.521E-09  &  0.866E-14   \\
\hline
\hline
\end{tabular}
\end{center}
\caption{MSD associated to the spatial integration
relative to our algorithm for various values of $N_c$,
the last row refers to usual pseudo-spectral code.
The data have been obtained for the one-dimensional CGLE
with time step $\tau =0.05$, periodic bounday conditions and for 
parameter values $c_1=3.5$ and $c_3=0.9$.
In this case the temporal MSD is $0.161E-02$.
The data of this Table have been previously reported
in \protect\cite{tor2}.}
\end{table}

\begin{table}[h]
\begin{center}
\begin{tabular}{c c c c}
\hline
\hline
$\Delta_x$ & 0.25 & 0.50 & 0.75\\
\hline
\hline
$N_c = 3$   &   & 0.510E-04  &   0.316E-03   \\
$N_c=5 $  &   & 0.463E-05   &  0.724E-04   \\
$N_c=10 $ & 0.654E-06   & 0.359E-06  & 0.264E-04   \\
$N_c=15 $ & 0.265E-07   & 0.308E-06   & 0.239E-04   \\
$N_c=20 $ & 0.113E-10  & 0.225E-06    & 0.724E-04   \\
FFT & 0.254E-13 & 0.214E-06  &  0.256E-04   \\
\hline
\hline
\end{tabular}
\end{center}
\caption{MSD associated to the spatial integration 
relative to our algorithm for various values of $N_c$,
the last row refers to usual pseudo-spectral code.
The data have been obtained for the two-dimensional CGLE
with time step $\tau =0.05$, periodic boundary conditions 
and for parameter values $c_1=3.5$ and $c_3=0.9$.
In this case the temporal MSD is $0.502E-03$.}
\end{table}

The CPU time required by the pseudo-spectral code will
increase with the linear dimension $N$ (where the total 
number of mesh points in $d$ dimension is $N^d$) as
$N^d \ln(N)$, instead our algorithm will scale as 
$ N^d f(N_c) $. In particular for the FHNE in the
one-dimensional case the ratio between the CPU time requested by the
code introduced here an that associated to a standard
FFT routine is $\gamma N_c / \ln(N)$, where $\gamma
\sim 0.15 $. Therefore for typical values ($N_c=10$,
$N=2048$) such ratio is $\sim 0.2$. In the two-dimensional case,
the ratio can be expressed as $\gamma [(N_c+1)(N_c+2)/2-1]/\ln(N)$
and for the FHNE we have $\gamma \sim 0.3$.
In this last situation for $N_c=4$ and $N=2048$,
the CPU time ratio has a value $\sim 0.6$.
Thus our algorithm is noticeably faster than usual
pseudo-spectral codes in one dimension, while in two dimension 
it becomes convenient
to use for grids larger than $N \sim 2048$.

\section{Conclusions\label{sec:Conclusions}}

In the present paper, an innovative
integration scheme for reaction-diffusion PDEs has
been reported. We have shown that such scheme 
is competitive, both in one and two dimensions,
with respect to usual time-splitting pseudo-spectral codes.
It combines high levels of accuracy with 
reduced cost in terms of CPU time and an extreme 
portability,
being applicable with a quite limited effort
to systems with different boundary conditions.
Future improvements of the present algorithm
should consist in determining an automatic procedure
for the choice of the $\alpha_h$-parameters.
We plan to implement this algorithm 
in a parallel enviroment and to extend its 
applicability also to more general classes of PDEs.

\nonumsection{Acknowledgements}
The first version of this integration 
scheme has been developed by one of us, for the one-dimensional CGLE, 
mainly following 
Peter Grassberger's ideas and in collaboration with Helge Frauenkron
\cite{tor1,tor2}.
We wish also to thank Markus B{\"a}r and Roberto Genesio for useful discussions.
A.T. acknowledges the hospitality of family Frese in Wuppertal (Germany) 
during the final write up of this text and Leonardo Torcini for providing
him a lap-top.

\nonumsection{References}

\end{document}